\documentstyle[12pt]{article}

\def\fnote#1#2{\begingroup\def\thefootnote{#1}\footnote{#2}
    \addtocounter{footnote}{-1}\endgroup}

\def\Email{makoto.natsuume@kek.jp}
\def\okamura{okamura@skyrose.phys.ocha.ac.jp}

\newcommand{\be}{\begin{equation}}
\newcommand{\ee}{\end{equation}}
\newcommand{\bea}{\begin{eqnarray}}
\newcommand{\eea}{\end{eqnarray}}

\newcommand{\rp}{r_+}
\newcommand{\rmi}{r_-}

\newcommand{\eq}[1]{(\ref{eq:#1})}

\newcommand{\half}{{1 \over 2}}

\def\rslash{\partial\kern-0.026em\raise0.17ex\llap{/}%
\kern0.026em\relax}
\def\Dslash{D\kern-0.15em\raise0.17ex\llap{/}\kern0.15em\relax}
\def\sqr#1#2{{\vcenter{\hrule height.#2pt
      \hbox{\vrule width.#2pt height#1pt \kern#1pt
          \vrule width.#2pt}
      \hrule height.#2pt}}}

\newcommand{\cH}{{\cal H}}
\newcommand{\cI}{{\cal I}}
\newcommand{\cJ}{{\cal J}}

\newcommand{\cM}{{\cal M}}
\newcommand{\cV}{{\cal V}}
\newcommand{\hH}{{\hat H}}
\newcommand{\back}[1]{ \stackrel{\circ}{#1} }
\newcommand{\og}{ {\back{g}} }
\newcommand{\oh}{ {\back{h}} }

\newcommand{\oG}{ {\back{G}} }
\newcommand{\oS}{ {\back{S}} }

\newcommand{\sGamma}{ { {}^{(n-1)}\!\Gamma } }
\newcommand{\sR}{ { {}^{(n-1)}\!R } }
\newcommand{\sG}{ { {}^{(n-1)}\!G } }
\newcommand{\tilg}{ {\tilde{g}} }
\newcommand{\tilh}{ {\tilde{h}} }
\newcommand{\tilp}{ {\tilde{p}} }

\newcommand{\tilR}{ {\tilde{R}} }
\newcommand{\tilU}{ {\tilde{U}} }
\newcommand{\tilV}{ {\tilde{V}} }
\newcommand{\tilW}{ {\tilde{W}} }

\newcommand{\tileta}{ {\tilde{\eta}} }
\newcommand{\tilpi}{ {\tilde{\pi}} }
\newcommand{\tilnabla}{ {\tilde{\nabla}} }
\begin{document}

\pagestyle{empty} 


\begin{flushright}
	KEK-TH-658, OCHA-PP-145 \\
	hep-th/9911062 \\
\end{flushright}

\vspace{18pt}
\begin{center}
{\large \bf Entropy for Asymptotically $AdS_3$ Black Holes}

\vspace{16pt}
Makoto Natsuume $^{1}$ \fnote{*}{\Email} and Takashi Okamura $^{2}$ \fnote{\dag}{\okamura}

\vspace{16pt}

{\sl $^{1}$ Theory Division\\
Institute of Particle and Nuclear Studies\\
KEK, High Energy Accelerator Research Organization\\
Tsukuba, Ibaraki, 305-0801 Japan}

\vspace{8pt}

{\sl $^{2}$ Department of Physics\\
Ochanomizu University\\
1-1 Otsuka, 2 Bunkyo-ku, Tokyo, 112-8610 Japan}

\vspace{12pt}
{\bf ABSTRACT}

\end{center}

\begin{minipage}{4.8in}
We propose that Strominger's method to derive the BTZ black hole entropy is in fact applicable to other asymptotically $AdS_3$ black holes and gives the correct functional form of entropies. We discuss various solutions in the Einstein-Maxwell theory, dilaton gravity, Einstein-scalar theories, and Einstein-Maxwell-dilaton theory. In some cases, solutions approach $AdS_3$ asymptotically, but their entropies do not have the form of Cardy's formula. However, it turns out that they are actually not ``asymptotically $AdS_3$" solutions. On the other hand, for truly asymptotically $AdS_3$ solutions, their entropies have the form of Cardy's formula. In this sense, all known solutions are consistent with our proposal. 
\end{minipage}


\vfill
\pagebreak

\pagestyle{plain}	
\setcounter{page}{1}	

\baselineskip=16pt

\section{Introduction}

Strominger has derived the Bekenstein-Hawking entropy of the Ba\~{n}ados-Teitelboim-Zanelli (BTZ) black hole \cite{BTZ,BHTZ} from the asymptotic Virasoro algebra \cite{andy}. Recently, we applied Strominger's method \cite{NOS} to a solution found by Mart\'{\i}nez and Zanelli (MZ) \cite{MZ}. This is the static solution of three-dimensional gravity with a conformal scalar field. The solution is not $AdS_3$ but it is asymptotically $AdS_3$ (See Section~\ref{sec-others} for the definition of ``asymptotically $AdS_3$"); therefore, it has the asymptotic Virasoro algebra. We found that the functional form of the Bekenstein-Hawking entropy agrees with the boundary conformal field theory (CFT) prediction. The overall numerical coefficient does not agree however; this is because this approach gives the ``maximum possible entropy" for the numerical coefficient \cite{carlip-9906}.

In this paper, we consider the extension of our approach. We propose that Strominger's method is in fact applicable to other asymptotically $AdS_3$ black holes and gives the correct functional form of entropies. In Section~\ref{sec-stationaryMZ}, we discuss the stationary black hole solution of three-dimensional gravity with a conformal scalar field. When angular momentum $J=0$, it reduces to the MZ solution. The Bekenstein-Hawking entropy formula of the black hole has the same form as Cardy's formula, {\it e.g.}, the entropy is the sum of the holomorphic part plus the anti-holomorphic part. 

In Section~\ref{sec-others}, we discuss other black holes which approach  $AdS_3$ asymptotically. We consider solutions of the Einstein-Maxwell theory, dilaton gravity, Einstein-scalar theories, and Einstein-Maxwell-dilaton theory \cite{BTZ}, \cite{KK-95}-\cite{KMN-97}. There have been known many solutions which approach $AdS_3$ asymptotically,
\footnote{When we write ``\ldots approaches $AdS_3$ asymptotically" in this paper, it means that the leading order terms of the metric are the same as the $AdS_3$ metric; this is different from ``asymptotically $AdS_3$" which constrains subleading terms in the metric (See Section~\ref{sec-others}).}
but whose entropies do not have the form of Cardy's formula. However, it turns out that they are actually not asymptotically $AdS_3$ solutions. On the other hand, for truly asymptotically $AdS_3$ solutions, their entropies have the form of Cardy's formula. In this sense, all known solutions are consistent with our proposal as far as we are aware.

In Appendix~\ref{sec:A}, we discuss the Hamiltonian formalism \cite{RT} to define the mass and angular momentum of black holes. We also prove that the charges in the Hamiltonian formalism are conformally invariant up to the reference terms for a generic Einstein-scalar theory in Appendix~\ref{sec:B}.

\section{The Stationary MZ Black Hole}\label{sec-stationaryMZ}

\subsection{The MZ Solution}

In this section, we consider the action given by
\be
S = \int_M d^{3}x \sqrt{-g} \left\{ \frac{R + 2l^{-2}}{16 \pi G} - \frac{1}{2} (\nabla \phi)^2 - \frac{1}{16} R \phi^2 \right\} + B'.
\label{eq:action}
\ee
The surface term $B'$ should be included so as to eliminate the second derivatives of the metric. The field equations are 
\be
R_{\mu \nu} - \frac{1}{2} g_{\mu \nu} R - \frac{1}{l^2} g_{\mu \nu} = 8 \pi G T_{\mu \nu},
	\label{eq:metricEOM}
\ee
\be
\nabla^2 \phi - \frac{1}{8} R \phi = 0,
	\label{eq:matterEOM}
\ee
where $T_{\mu \nu}$ is the matter stress tensor given by
\be
T_{\mu \nu} = \nabla_{\mu} \phi \nabla_{\nu} \phi - \frac{1}{2} g_{\mu \nu} (\nabla \phi)^2 
	+ \frac{1}{8} \{ g_{\mu \nu} \nabla^2 - \nabla_{\mu} \nabla_{\nu} + R_{\mu \nu} - \frac{1}{2} g_{\mu \nu} R \} \phi^2.
\label{eq:em}
\ee
The matter stress tensor \eq{em} is traceless. Therefore,
\be
R = -\frac{6}{l^2}.
\ee
This immediately implies that the solutions to Eqs.~\eq{metricEOM} and \eq{matterEOM} are asymptotically $AdS_3$.

The MZ solution is 
\bea
ds^2 & = & - \frac{1}{l^2} \left(\rho + \frac{\rp}{2}\right)^2 \left(1 - \frac{\rp}{\rho}\right) d\hat{t}^2 
	+ \frac{l^2 d\rho^2}{(\rho + \frac{\rp}{2})^2(1 - \frac{\rp}{\rho})} + \rho^2 d \hat{\varphi}^2,
		\label{eq:mz} \\
\phi^2 & = & \frac{\rp}{\pi G (2\rho + \rp)}.
\eea
The coordinates take $ - \infty < \hat{t} < + \infty $, $ 0 \leq \hat{\varphi} < 2 \pi $, and $ 0 \leq \rho < + \infty $. See refs.~\cite{NOS,MZ} for further properties of the solution.

\subsection{Stationary Solution}

A stationary metric can be written as
\be
ds^2 = - (N^\perp)^2(r) dt^2 + f^{-2}(r) dr^2 + R^2(r) \{ d\varphi + N^\varphi(r) dt \}^2.
\label{eq:stationary}
\ee
In 2+1 dimensions, a stationary black hole solution can be generated from a static black hole solution by a ``coordinate transformation" since the transformation can be done only locally \cite{stachel}. 

This is because the spacetime is not simply connected; closed curves which encircle the horizon cannot be shrunk to a point. Thus, the first Betti number $b_1$ is equal to one and there is a closed 1-form which is not exact. Suppose a stationary solution is generated from the following transformation:
\be
\hat{t} = a t + b \varphi,
\ee
where $\hat{x}^{\mu}$ and $x^{\mu}$ are the coordinates of a static and a stationary solution; $a$ and $b$ are constants. There is a timelike Killing vector $\eta=\partial_t$ in both solutions. The vector fields are mapped onto each other under a global diffeomorphism. Given the vector, one can construct a 1-form $V$ which is inverse to $\eta$; $V_{\mu} \equiv \eta_{\mu}/|\eta|^2$. In the static solution, $V=d\hat{t}$, which is an exact 1-form. This is transformed as $V=a dt +b d\varphi$ formally in the stationary solution. This is a closed 1-form but not exact because the period of $V$ about closed curves $C$ enclosing the horizon does not vanish:
\be
\int_{C} V = b \int_{C} d\varphi = 2 \pi b.
\ee
Thus, these 1-forms are mapped to each other only {\it locally}. Because a metric maps vectors to 1-forms, this means that metrics cannot be mapped to each other either.

One also has to make the transformation $\hat{\varphi} = b t + a \varphi$ and $\rho^2 \sim r^2/(a^2-b^2)$ as $r \rightarrow \infty$ in order that the resulting metrics become asymptotically $AdS_3$. We therefore make the following coordinate transformation to the $\rp=1$ solution:
\bea
\left(\begin{array}{c}\hat{t} \\ l \hat{\varphi} \end{array}\right) &=&
	\frac{1}{l} \left(\begin{array}{cc} \rp & -\rmi \\ -\rmi & \rp \end{array}\right)
	\left(\begin{array}{c} t \\ l \varphi \end{array}\right), 
\label{eq:coord1} \\
\rho^2 &=& l^2 \frac{r^2 - \rmi^2}{\rp^2 - \rmi^2}.
\label{eq:coord2}
\eea
This is the same coordinate transformation used to generate the stationary BTZ black hole. The coordinate transformation ensures that $N^\perp$ and $g^{rr}$ vanish at $r=r_\pm$ under appropriate conditions.

Applying the coordinate transformation, one obtains
\bea
ds^2 &=& -\frac{1}{l^2} \frac{ (r^2-\rmi^2) \{4(r^2-\rmi^2)-(\rp^2-\rmi^2)(3+\frac{l}{\rho})\} }
		{4r^2 - \rmi^2(1-\frac{l}{\rho})} dt^2 \nonumber \\
	&& \qquad + \frac{4l^2 r^2 dr^2}{ (r^2-\rmi^2) \{4(r^2-\rmi^2)-(\rp^2-\rmi^2)(3+\frac{l}{\rho})\} } \nonumber \\
	&& \qquad + \{ r^2 - \frac{1}{4} \rmi^2 (1-\frac{l}{\rho}) \}
	\left\{ d\varphi - \frac{1}{l} \frac{\rp\rmi(3+\frac{l}{\rho})}{4r^2 - \rmi^2(1-\frac{l}{\rho})} dt \right\}^2, 
\label{eq:stationaryMZ} \\
\phi^2 &=& \frac{1}{\pi G} \frac{\sqrt{\rp^2-\rmi^2}}{2\sqrt{r^2-\rmi^2} + \sqrt{\rp^2-\rmi^2}}.
\eea
As in the BTZ black hole, we now identify $ \varphi \approx \varphi + 2 \pi $ instead of $\hat{\varphi}$. 

Since $N^\perp$ and $g^{rr}$ vanish at $r=r_\pm$, $r_\pm$ represent the locations of the outer and the inner horizon. This can be seen by using Eddington-Finkelstein like coordinates:
\be
dv = dt + h(r) dr, \qquad d\tilde{\varphi} = d\varphi + k(r) dr,
\ee
where
\bea
h(r) &=& - \frac{1}{g_{00}} \left\{ 1 + \frac{R N^\varphi}{N^\perp} \sqrt{1+\frac{g_{00}}{f^2}} \right\}, \\
k(r) &=& \frac{\sqrt{f^2+g_{00}}}{f N^\perp R}.
\eea
Then, the metric becomes
\bea
ds^2 &=& - (N^\perp)^2 dv^2 + 2 dv dr + R^2 (d\tilde{\varphi} + N^\varphi dv)^2 \nonumber \\
	&& \qquad\qquad + \frac{R}{g_{00}} \left\{ R N^\varphi + N^\perp \sqrt{1+\frac{g_{00}}{f^2}} \right\} 2d\tilde{\varphi} dr.
\eea
Thus, the surface $r=\rp$ is a null surface generated by the geodesics
\be
r(\lambda) = \rp, \qquad \frac{d\tilde{\varphi}}{d\lambda} + N^\varphi(\rp) \frac{dv}{d\lambda} = 0.
\label{eq:geodesics}
\ee
Also, the surface is a marginally trapped surface since any null geodesics satisfy
\be
\frac{dv}{d\lambda} \frac{dr}{d\lambda} = 
-\frac{\rp^2}{2} \left( \frac{d\tilde{\varphi}}{d\lambda} + N^\varphi(\rp) \frac{dv}{d\lambda} \right)^2 \leq 0
\ee
at $r=\rp$. Here, we used $g_{\tilde{\varphi} r} = 0$ at $r=\rp$. So, $r$ decreases or remains constant [for the geodesics \eq{geodesics}] as $v$ increases.

On the other hand, the ``inner horizon" $r=\rmi$ is actually singular because
\be
R^{\mu\nu\rho\sigma} R_{\mu\nu\rho\sigma} = \frac{12 \{ 32(r^2-\rmi^2)^3 + (\rp^2-\rmi^2)^3 \}}{32 l^4 (r^2-\rmi^2)^3}.
\ee

We use the Hamiltonian formalism \cite{RT} to define the mass and angular
momentum of the black hole. In Appendix \ref{sec:A}, 
we discuss the canonical formalism for a slightly more general action 
than Eq.~\eq{action}. For the action \eq{action}, the charges are given by
Eq.(\ref{eq:cJforpure}).
Using the zero-mass black hole
\be
ds^2=\og_{\mu\nu} dx^\mu dx^\nu
= -{r^2 \over l^2} dt^2 + {l^2 \over r^2} dr^2 
+ r^2 d\varphi^2
\ee
as the reference spacetime, one gets
\bea
8G M &=& \frac{3}{4l^2} (\rp^2 + \rmi^2), \label{eq:mass} \\
8G J &=& \frac{3}{2l} \rp \rmi, \label{eq:angular}
\eea
for the stationary solution \eq{stationaryMZ}.

In order for the solution to be a black hole ({\it i.e.}, no naked singularity), one must have $\rp \geq \rmi$
or
\be
M>0, \qquad\qquad |J| \leq Ml.
\ee
However, in the extreme limit $|J| = Ml$, the singularity completely disappears; the solution becomes the extreme BTZ black hole by an appropriate coordinate transformation.

The static limit $r=r_{erg}$ is always outside the outer horizon since $g_{00}$ is a monotonically decreasing function and $g_{00}>0$ at the outer horizon.

The metric \eq{stationaryMZ} approaches $AdS_3$ asymptotically,
{\it i.e.},
\be
ds^2 \rightarrow - \frac{r^2}{l^2} dt^2 + \frac{l^2}{r^2} dr^2 + r^2 d\varphi^2 .
\ee
Moreover, the metric satisfies the boundary conditions given by Brown and Henneaux \cite{BH}. Therefore, there should exist the asymptotic Virasoro algebra.

\subsection{Thermodynamics}

The entropy of the black hole is given by
\be
S_{BH} = \frac{\pi \rp}{3G}.
\label{eq:entropy1}
\ee
The entropy does not satisfy the area law $S_{BH}=A/(4G)$. This is because the solution \eq{stationaryMZ} is not written in terms of the Einstein metric \cite{NOS}.

The horizon $r=\rp$ is also a Killing horizon, {\it i.e.}, a null surface to which a Killing vector is normal. The relevant Killing vector is
\be
\chi = \frac{\partial}{\partial t} + \Omega_H \frac{\partial}{\partial \varphi},
\ee
where $\Omega_H$ is the angular velocity at the horizon:
\be
\Omega_H = \left. \frac{d\varphi}{dt}\right|_{\rp} 
	= - \left. \frac{g_{t\varphi}}{g_{\varphi\varphi}}\right|_{\rp} = \frac{\rmi}{\rp l}.
\ee
The surface gravity $\kappa$ is defined by the Killing vector as
\be
\kappa^2 = -\frac{1}{2} (\nabla^{\mu} \chi^{\nu}) (\nabla_{\mu} \chi_{\nu}).
\ee
Therefore, we define the Hawking temperature as
\be
T = \frac{\kappa}{2\pi} = \frac{9}{16\pi^2} \frac{\rp^2 - \rmi^2}{\rp l^2}.
\label{eq:temp}
\ee
The Hawking temperature depends on the normalization of the timelike Killing vector. Since the spacetime is asymptotically $AdS_3$, there is no natural normalization we can take. Equation~\eq{temp} is the Hawking temperature with respect to the Killing vector $\partial_t$. But the ``Killing time" is not the same as the ``affine time" of the asymptotic observer unlike asymptotic flat spacetimes; therefore, the physical significance of the ``Killing time" is not entirely clear. The temperature $T(R)$ as seen by the observer at distance $r=R$ is given by the Tolman relation \cite{wald}:
\be
T(R) = \frac{T}{N^\perp(R)},
\ee
which vanishes at infinity; there is an infinite redshift at infinity.

From Eqs. \eq{mass}, \eq{angular}, \eq{entropy1}, and \eq{temp}, one can check that the first law of thermodynamics is satisfied:
\be
dM = T dS + \Omega_H dJ.
\ee

Now, in terms of the mass \eq{mass} and the angular momentum \eq{angular}, the entropy is given by
\be
S_{BH} = \frac{2\pi}{3} \left\{ \sqrt{\frac{2l(l M+J)}{3G}} + \sqrt{\frac{2l(l M-J)}{3G}} \right\}.
\label{eq:entropy2}
\ee
This has the same form as Cardy's formula, {\it i.e.}, the entropy is the sum of the holomorphic part plus the anti-holomorphic part. Moreover, $M$ and $J$ appear in the same form as $L_0$ and $\bar{L}_0$ eigenvalues \cite{NOS}. All these may be reminiscent of the asymptotic CFT. 

However, the overall numerical coefficient does not agree with the asymptotic CFT's result. The central charge does not change from the pure-gravity result due to the no-hair theorems \cite{NOS}. Then, one get for the CFT prediction as
\be
S_{asymp} = \pi \left\{ \sqrt{\frac{l(l M+J)}{2G}} + \sqrt{\frac{l(l M-J)}{2G}} \right\}.
\ee

\section{Other $AdS_3$ Solutions}\label{sec-others}

In this section, we briefly discuss other asymptotically $AdS_3$ solutions for three-dimensional gravity with a cosmological constant and matter fields. Actually, only known asymptotically $AdS_3$ solutions are those for Einstein-scalar theories. Thus, these are the only solutions to which one can apply Strominger's method. This is fortunate however; entropies for other solutions does not have the same form as Cardy's formula. In this sense, all known solutions are consistent with our proposal.

To make discussion clear, we give the definition of ``asymptotically $AdS_3$":
\begin{enumerate}
\item[(i)]  They should contain the black hole solution in question.
\item[(ii)]  They should be invariant under the $AdS_3$ group $O(2,2)$ at spatial infinity.
\item[(iii)]  They should make the surface integrals associated with the generators of $O(2,2)$ finite.
\end{enumerate}
In particular, conditions~(i) and (ii) require that 
\begin{enumerate}
\item[(i')]  The metric of the solution should satisfy the boundary conditions given by Brown and Henneaux \cite{BH}.
\end{enumerate}

\subsection{Einstein-Maxwell Theory}

A static electrically charged solution was obtained by \cite{BTZ}. It was later generalized by various authors \cite{KK-95,relatedtoKK,HW-95,chen-98}. The most general solution in this category is \cite{chen-98}. The solution found by \cite{HW-95} is not a black hole, so is uninteresting for our purpose. These solutions approach $AdS_3$ asymptotically, but none of the solutions are asymptotically $AdS_3$; the gauge field produces $O(\ln r)$ terms in metrics. Note that condition~(i') constrains the subleading terms in metrics as well. Moreover, the mass and angular momentum diverge because of the logarithmic term, thus violating condition~(iii) as well.

\subsection{Dilaton Gravity}

Solutions are found by \cite{CM-94,lemos-94,SKL-95,CM-95,LZ-95,SL-96}. The most general solution in this category is \cite{CM-95,SL-96}. The action considered by \cite{SKL-95,SL-96} is (in the ``string metric")
\be
S = \frac{1}{16\pi G} \int d^{3}x \sqrt{-g}\, e^{-2\phi} \left\{ R - 4\omega (\nabla \phi)^2 + \frac{2}{l^2} \right\}.
\ee
The static solutions in the string metric are given by
\bea
ds^2 &=& - a^2 \left\{ r^2 - r_+^2(\frac{r_+}{r})^{\frac{1}{\omega+1}} \right\} dt^2 
	+ \frac{dr^2}{a^2 \left\{ r^2 - r_+^2(\frac{r_+}{r})^{\frac{1}{\omega+1}} \right\}} 
	+ r^2 d\varphi^2, \\
e^{2\phi} &=& (ar)^{-\frac{1}{\omega+1}},
\eea
where
\be
a=\frac{\sqrt{2}(\omega+1)}{\sqrt{(\omega+2)(2\omega+3)}} \frac{1}{l}, \qquad
8GM=\frac{\omega+2}{\omega+1} (a r_+)^{2+\frac{1}{\omega+1}}. 
\ee
When $\omega=-1$, the action corresponds to the low-energy bosonic string action with a world-sheet conformal anomaly. The $\omega=-1$ solution is not asymptotically $AdS_3$ though. The solutions satisfy condition~(i') if $\omega>-1$. The action of Chan and Mann \cite{CM-94,CM-95} is in the Einstein metric and solutions are not asymptotically $AdS_3$ except for pure gravity. However, the solutions satisfy condition~(i') in the ``string metric." The solutions after the transformation become those of \cite{SKL-95,SL-96}; they satisfy condition~(i') when $1<N<2$ in their notation. Refs.~\cite{lemos-94,LZ-95} consider the $d=3$ Kaluza-Klein theory with a cosmological constant and corresponds to the $\omega=0$ action.

For the static solutions, the entropy is written as
\be
S_{BH} = \frac{\pi}{2aG} \left\{ \frac{\omega+1}{\omega+2} 8GM \right\}^{\frac{1}{2}+\frac{1}{2(2\omega+3)}}.
\ee
Thus, the entropy does not have the form of Cardy's formula except when $\omega \rightarrow \infty$ (pure gravity). 

However, one cannot apply Strominger's method to the solution. One actually needs to enforce stronger conditions than Brown-Henneaux's boundary conditions. Otherwise, the charges $\cJ[\eta]$ associated with the symmetries diverge in general for deformed spacetimes. In other words, they do not satisfy condition~(iii).

For instance, consider the second term of Eq.~\eq{cJforpure}: for metrics which satisfy condition~(i'),
\be
G^{ijkl} \left( \eta^\perp D_j q_{kl} - q_{kl} D_j \eta^\perp \right) \sim O(1).
\ee
Since $U \propto \exp(-2\phi) \propto r^{1/(\omega + 1)}$, the mass is generally expected to diverge. Thus, the boundary conditions one really needs are ($a,b=t,\varphi$)
\renewcommand{\arraystretch}{1.5} 
\be 
\begin{array}{ll}
	e^{-2\phi}q_{rr}=O(r^{-4}), & e^{-2\phi}q_{ra}=O(r^{-3}), \\
	e^{-2\phi}q_{ab}=O(1), & e^{-2\phi}=O(r^{1/(\omega+1)}), 
\end{array} 
\ee 
\renewcommand{\arraystretch}{1}%
for the metric perturbation $q_{\mu\nu} = g_{\mu\nu} - \og_{\mu\nu}$, where $\og_{\mu\nu}$ is the zero-mass black hole metric. Such a condition restricts the Virasoro algebra so that the canonical realization of the Virasoro algebra does not exist.

\subsection{Einstein-Scalar Theories}

Chan has obtained a variety of solutions \cite{chan-96} which include the MZ solution as a special case.
\footnote{We use the word ``dilaton" only if a theory is invariant under the dilatation, $\phi \rightarrow \phi+a$. When gauge fields are present, the gauge fields are allowed to transform as well. }
Some of them are asymptotically $AdS_3$. In the following, $-\Lambda<0$ is an effective cosmological constant at infinity, $M$ is the mass computed in the Hamiltonian formalism, and we use the unit $8G=1$ for simplicity.
\begin{itemize}
\item Solutions~(33) and (40) of ref.~\cite{chan-96} are given by
\bea
ds_{(33)}^2 &=& - \left\{ \Lambda r^2 - M(1-\frac{\sqrt{M}}{br}) \right\} dt^2 
	+ \frac{dr^2}{\Lambda r^2 - M(1-\frac{\sqrt{M}}{br})} 
	+ r^2 d\varphi^2, \nonumber \\
\phi_{(33)} &=& \frac{r}{r-\frac{3\sqrt{M}}{2br}}, \nonumber
\eea
%
%
\bea
ds_{(40)}^2 &=& - \left\{ \Lambda r^2 - M(1-\frac{M}{\lambda r^2}) \right\} dt^2 
	+ \frac{dr^2}{\Lambda r^2 - M(1-\frac{M}{\lambda r^2})} 
	+ r^2 d\varphi^2, \nonumber \\
\phi_{(40)} &=& \frac{r^2}{r^2-\frac{2M}{\lambda}}, \nonumber
\eea
%
%
respectively. Here, $b$ ($=\sqrt{M}/B$ in \cite{chan-96}) and $\lambda$ ($=M/L$ in \cite{chan-96}) are coupling constants appeared in actions with appropriate dimensions. For a particular value of $B$, the solution~(33) becomes the MZ solution. 

The solutions satisfy condition~(i'). Using Eqs.~\eq{deltaH} and \eq{cI}, one can show that scalars do not contribute to the charges $\cJ[\eta]$ under asymptotic behavior of canonical variables and surface deformation vector; we have the same surface term as the one for pure gravity. Thus, the charges $\cJ[\eta]$ obey the Virasoro algebra and the central charge does not change from the pure-gravity result. 

These black holes have entropies $S_{BH} \propto \sqrt{M}$, so the functional forms agree with CFT predictions. 

\item Solutions~(22) and (28): solution~(22) is given by
\bea
ds_{(22)}^2 &=& - (\Lambda r \sqrt{r^2-B^2}) dt^2 
	+ \frac{r dr^2}{\Lambda r (r^2-B^2)^{\frac{3}{2}}}
	+ r^2 d\varphi^2, \nonumber \\
\phi_{(22)} &=& \frac{i}{4} \arccos (1-\frac{2B^2}{r^2}). \nonumber
%
\eea
Here, $B$ is an integration constant related to the mass: $M=\frac{1}{2} \Lambda B^2$ in the Hamiltonian formalism. Note that the mass differs by a numerical factor from the one given by Chan (in his quasilocal stress tensor formalism). The scalar $\phi_{(22)}$ is imaginary; this changes the sign of the kinetic term, but the action remains hermitian. Chan considers the negative-mass case $B^2<0$. Then, the solution is not a black hole, but rather a naked singularity located at $r=0$. 

Solution~(28) is given by
\bea
ds_{(28)}^2 &=& - \left\{ \Lambda r^2 - M(1-\frac{M}{\lambda r^2}) \right\} dt^2 
	+ \frac{(1-\frac{2M}{\lambda r^2})^2}{\Lambda r^2 - M(1-\frac{M}{\lambda r^2})} dr^2
	+ r^2 d\varphi^2, \nonumber \\
\phi_{(28)} &=& \frac{1}{\sqrt{2}}  \arccos (\sqrt{\frac{2M}{b}} \frac{1}{r}). \nonumber
%
\eea
Here, $\lambda$ ($=M/L$ in \cite{chan-96}) is a coupling constant appeared in the action with an appropriate dimension. 

The solutions satisfy condition~(i'). These black holes have entropies $S_{BH} \propto \sqrt{M}$. However, scalars contribute to the charges $\cJ[\eta]$, so one needs to reexamine the algebra in order to compare numerical coefficients. 

\item Solutions~(24) and (38) are given by
\bea
ds_{(24)}^2 &=& - \left\{ \Lambda r^2 - M(1-\frac{\sqrt{M}}{br}) \right\} dt^2 
	+ \frac{(1-\frac{3\sqrt{M}}{2br})^2}{\Lambda r^2 - M(1-\frac{\sqrt{M}}{br})} dr^2
	+ r^2 d\varphi^2, \nonumber \\
\phi_{(24)} &=&  \arccos (\sqrt{\frac{3}{2br}} M^{\frac{1}{4}}), \nonumber
%
\eea
\bea
ds_{(38)}^2 &=& - \left\{ \Lambda r^2 + A(1+kr) \right\} dt^2 
	+ \frac{dr^2}{\Lambda r^2 + A(1+kr)} dr^2
	+ r^2 d\varphi^2, \nonumber \\
\phi_{(38)} &=& \frac{2}{2+kr}, \nonumber
\eea
respectively. Here, $b$ ($=\sqrt{M}/B$ in \cite{chan-96}), $A$ and $k$ are coupling constants appeared in actions with appropriate dimensions. The solutions do not satisfy condition~(i'), so are not asymptotically $AdS_3$.
\end{itemize}

Incidentally, some of Chan's actions contain $M$ explicitly as a coupling constant, so $M$ is not an integration constant. One can redefine the coupling constants so that $M$ does not appear in actions ({\it e.g.}, for solution~(33), define $b$ such that $b=\sqrt{M}/B$). He argues that this causes a problem since the mass diverges in his quasilocal stress tensor formalism. However, the mass remains finite in the Hamiltonian formalism and agrees with $M$ in the original actions. 

\subsection{Einstein-Maxwell-Dilaton Theory}

Solutions are obtained by \cite{chen-98,CM-94,KP-96,KMN-97}. The action of \cite{chen-98,CM-94,KP-96} is
\be
S = \frac{1}{16\pi G} \int d^{3}x \sqrt{-g}\, e^{-2\phi} \left\{ R - 4\omega (\nabla \phi)^2 + \frac{2}{l^2} - F^2 \right\}
\ee
in the string metric. The gauge field $F_{\mu\nu}$ couples to the dilaton in the NS-NS gauge field form. The most general solution for the action is \cite{chen-98}. 

The static electric solutions in the string metric are given by
\bea
ds^2 &=& - U(r) dt^2 + (\frac{\omega+3}{\omega+1})^2 \frac{dr^2}{U(r)} + r^2 d\varphi^2, \\
U(r) &=& \frac{2(\omega+3)^2}{(\omega+2)(2\omega+3)} (\frac{r}{l})^2 
	- m (\frac{l}{r})^{\frac{1}{\omega+1}} + \frac{2}{\omega+2} q^2 (\frac{l}{r})^{\frac{2}{\omega+1}}, \\
A_0 &=& -q (\frac{l}{r})^{\frac{1}{\omega+1}}, \\
e^{2\phi} &=& (\frac{l}{r})^{\frac{1}{\omega+1}},
\eea
where $m$ is a parameter related to the mass of the solution. The solutions of \cite{chen-98,CM-94} satisfy condition~(i') in the string metric when $\omega>-1$. However, it has the same problem as dilaton gravity. However, in general, the entropy does not have the same form as Cardy's formula either. Ref.~\cite{KMN-97} considers a gauge field which does not couple to the dilaton in the NS-NS field form. The solution of \cite{KMN-97} does not approach $AdS_3$ asymptotically. Incidentally, the presence of gauge fields could extend the Virasoro algebra.

\vspace{.1in}
\begin{center}
    {\Large {\bf Acknowledgements} }
\end{center}
\vspace{.1in} 

We would like to thank A. Hosoya, N. Ishibashi, T. Maki, J. Polchinski, T. Sakai, and M. Sato for useful discussions. The work of M.N. was supported in part by the Grant-in-Aid for Scientific Research (11740161) from the Ministry of Education, Science and Culture, Japan.

\appendix
\section{Charges in the Regge-Teitelboim Formalism} \label{sec:A}
We consider the following action for a scalar-tensor theory in $n$-dimensional spacetime,
\begin{eqnarray}
	& &S = {1 \over 16\pi G} \int d^n x \sqrt{-g} \left\{ 
	U(\phi) R + 2 l^{-2} \right\} + S_M,
\label{eq:generalaction} \\
	& &S_M = \int d^n x \sqrt{-g} \left\{
	-\half W(\phi) \left( \nabla \phi \right)^2 - V(\phi) \right\}.
\label{eq:generalmatter}
\end{eqnarray}
According to the standard ADM decomposition, the bulk Hamiltonian is written in the form
\begin{equation}
	H[ N ] = \int_\Sigma d^{n-1} x 
	\left[ N^\perp \cH_\perp + N^i \cH_i \right],
\label{eq:bulkHam}
\end{equation}
where $N^\perp$, $N^i$ and $\Sigma$ are the lapse, the shift functions and a time slice, respectively. Here, we have dropped the possible surface terms. The Hamiltonian and momentum constraints are given by
\begin{eqnarray}
	& &\cH_\perp = {1 \over 2 \sqrt{h}} \pmatrix{ \pi^{ij} & p \cr}
	\cM^{-1} \pmatrix{ \pi^{kl} \cr p \cr} + \sqrt{h} \cV,
\label{eq:bulkHamconst} \\
	& &\cH_i = -2 \sqrt{h} D_j \left( 
	{\pi_i{}^j \over \sqrt{h}} \right) + p D_i \phi,
\label{eq:bulkmomconst}
\end{eqnarray}
where $\pi^{ij}$ and $p$ are the canonical conjugate momentum of $h_{ij}$ and $\phi$ respectively, and
\begin{eqnarray}
	& &\cM^{-1} = \pmatrix{ {32 \pi G \over U} G_{ijkl} 
	+ {2 U' \over (n-2) U}\beta h_{ij} h_{kl} & -\beta h_{ij} \cr
	-\beta h_{kl} & {n-2 \over 2} {U \over U'} \beta \cr },
\\
	& &G_{ijkl} = h_{i(k}h_{l)j}-{1 \over n-2}h_{ij} h_{kl},
\hspace{1cm}
	\beta = { U' \over {n-2 \over 2} UW+{U'^2 (n-1) \over 16\pi G} },
\\
	& &\cV = -{1 \over 16\pi G} \left( U~ \sR + 2 l^{-2} \right)
	+\half W \left( D\phi \right)^2 + V + {1 \over 8\pi G} \Delta U.
\end{eqnarray}
Here, $U' = \frac{\partial U}{\partial \phi}$, ${}^{(n-1)}\!R$ is the $(n-1)$-dimensional Ricci scalar, and $D_{i}$ is the covariant derivative with respect to $h_{ij}$.

The Regge-Teitelboim formalism starts with constructing the Hamiltonian, which has well-defined functional derivatives with respect to canonical variables for any surface deformation vector.

We list some convenient formulae:
\begin{eqnarray}
	& &\delta \left( \sGamma{}^i_{jk} \right)
	=\left[ h^{in} \delta^l_{(j} \delta^m_{k)}
	-\half h^{il} \delta^m_j \delta^n_k \right]~D_l \delta h_{mn},
\\
	& &\delta \left( \sqrt{h}~ \sR \right) = 
	- \sqrt{h}~ \sG{}^{ij} \delta h_{ij}
	+ \sqrt{h} D_i \left( G^{ijkl} D_j \delta h_{kl} \right),
\\
	& &\delta \left( \sqrt{h}~ h^{ij} \right) = 
	- \sqrt{h}~ S^{ijkl} \delta h_{kl},
\\
	& &S^{ijkl}= h^{i(k} h^{l)j} -\half h^{ij} h^{kl}
	=G^{ijkl}+\half h^{ij} h^{kl},
\\
	& &\delta \left( \sqrt{h} \Delta U \right) = 
	- \sqrt{h}~D_i \left[ S^{ijkl}~\delta h_{kl}~D_j U \right].
\end{eqnarray}

Using the above formulae, one can obtain the variation of the bulk Hamiltonian with a surface deformation vector $\eta = \eta^\perp n + \eta^i \partial_i$:
\begin{eqnarray}
	\delta H [ \eta ] &=& (\mbox{bulk terms}) 
	- \oint_{\partial \Sigma} dS_i~ \cI^i [ \eta ],
\label{eq:deltaH}
\\
	\cI^i [ \eta ] &=& 
	2 { \delta\left( \eta_j \pi^{ij} \right) \over \sqrt{h} }
	- \eta^i { \pi^{jk} \delta h_{jk} + p \delta\phi \over \sqrt{h} }
\nonumber \\
	& &+ {1 \over 16 \pi G} \left[ U G^{ijkl} \left( 
	\eta^\perp D_j \delta h_{kl} 
	- \delta h_{kl} D_j \eta^\perp \right)
	-\eta^\perp \delta(h^{ij}) D_j U \right]
\label{eq:cI} \\
	& & - {1 \over 8 \pi G} \left[ 
	\left( \eta^\perp \right)^2 e^{-\gamma} 
	D^i \left( e^\gamma {\delta U \over \eta^\perp} \right) \right],
\label{eq:variationHamiltonian} \nonumber 
\end{eqnarray}
where
\begin{equation}
	\gamma = 8 \pi G \int d\phi {W(\phi) \over U'(\phi)}.
\label{eq:gammadef}
\end{equation}
The bulk terms give the equations of motion. Note that we have not yet considered any fall-off conditions in Eq.~\eq{cI}.

If one can rewrite the surface term \eq{deltaH} in terms of a total variation by some asymptotic conditions, $\oint_{\partial \Sigma} dS_i \, \cI^i [ \eta ] =\delta \cJ [ \eta ]$, the Hamiltonian $\hH[\eta]=H[\eta]+\cJ[\eta]$ has well-defined variational derivatives. For example, for three-dimensional Einstein gravity (with or without a conformally coupled scalar field), $\cJ$ is given by
\begin{equation}
	\cJ [ \eta ] = \oint_{\partial \Sigma} d \oS_i
	\left[ 2 { \left( \eta_j \pi^{ij} \right) \over \sqrt{\oh} }
	+ {1 \over 16 \pi G} \left\{ \oG{}^{ijkl} \left( 
	\eta^\perp D_j q_{kl} 
	- q_{kl} D_j \eta^\perp \right) \right\} \right]
\label{eq:cJforpure}
\end{equation}
under the asymptotically $AdS_3$ conditions \cite{NOS}. Here, $q_{\mu\nu}=g_{\mu\nu}-\og_{\mu\nu}$ and the reference metric $\og_{\mu\nu}$ is a metric satisfying the asymptotically $AdS_3$ conditions, which is usually chosen as the globally $AdS_3$ or the zero-mass black hole.

The Hamiltonian $\hH[\eta]$ is the generator of the transformation associated with the surface deformation vector $\eta$. If the vector $\eta$ is an asymptotic Killing vector, then $\hH[\eta]$ is the generator of the asymptotic symmetry and gives a conserved charge associated with it. Since we have the constraint $H[\eta] \approx 0$, the numerical value of the $\hH[\eta]$ is just the surface term $\cJ[\eta]$.

If there is a timelike or rotational asymptotic Killing vector, $\partial/\partial t$ or $\partial/\partial \varphi$, one can define the mass or angular momentum as $\cJ[\partial/\partial t]$ and $\cJ[\partial/\partial \varphi]$ respectively.

It is obvious from the above discussion that in the Regge-Teitelboim formalism, we can fix the values of the charges only up to the reference values.

\section{Conformal Invariance of Charges} \label{sec:B}

In this appendix, we consider the dependence of the charges on the choice of the conformal frame.

A conformal transformation such as $\tilg_{\mu\nu}=\Omega^2(\phi)~ g_{\mu\nu}$ is merely a change of variables, so physical quantities such as the mass and angular momentum should be invariant under a conformal transformation. We will show that the charges in the Regge-Teitelboim formalism are actually invariant under a conformal transformation, up to the reference values.

Under a conformal transformation, various geometrical quantities transform as
\begin{eqnarray}
	\tilg^{\mu\nu} &=& \Omega^{-2}~ g^{\mu\nu},
\\
	\sqrt{- \tilg} &=& \Omega^n \sqrt{- g},
\\
	R &=& \Omega^2 \left[ \tilR + 2(n-1) \tilde \sqr73 \ln \Omega
	-(n-2)(n-1) \left( \tilnabla \ln \Omega \right)^2 \right].
\end{eqnarray}
Therefore we can rewrite the action (\ref{eq:generalmatter}) as
\begin{eqnarray}
	S &=& {1 \over 16\pi G} \int d^n x \sqrt{-\tilg} \left\{ 
	\tilU(\phi) \tilR + 2 l^{-2} \right\}
\nonumber \\
	& &+ \int d^n x \sqrt{-\tilg} \left\{
	-\half \tilW(\phi) \left( \tilnabla \phi \right)^2 - \tilV(\phi) 
	\right\} 
\\
	& & + {n-1 \over 8 \pi G} \int d^n x \sqrt{-\tilg}~ \tilnabla
	\left( \tilU \tilnabla \ln \Omega \right),
\nonumber
\end{eqnarray}
where 
\begin{eqnarray}
	\tilU(\phi) &=& \Omega^{2-n}(\phi) U(\phi),
\\
	\tilW(\phi) &=& \Omega^{2-n}(\phi) \left[ W(\phi)
	+{n-1 \over 8 \pi G} U(\phi) \left({d \over d\phi} \ln \Omega \right)
	{d \over d\phi} \ln ( U^2 \Omega^{2-n} ) \right],
\\
	\tilV(\phi) &=& \Omega^{-n}(\phi) V(\phi)
	+{l^{-2} \over 8 \pi G} \left( 1- \Omega^{-n} \right).
\end{eqnarray}

Since the bulk Hamiltonian is given by neglecting surface term, the bulk Hamiltonian in the {\it tilde} frame is simply given by replacing various quantities in the original Hamiltonian with the corresponding quantities in the {\it tilde} frame.

The canonical variables transform as $(h_{ij}, \phi~;~ \pi^{ij}, p) \rightarrow (\tilh_{ij}, \tilde\phi=\phi~;~ \tilpi^{ij}, \tilp)$, where
\begin{eqnarray}
	\tilh_{ij} &=& \Omega^2(\phi) h_{ij},
\\
	\tilpi^{ij} &=& \Omega^{-2}(\phi) \pi^{ij},
\\
	\tilp &=& p - 2 \left({d \over d\phi} \ln \Omega \right) \pi.
\end{eqnarray}

After some calculation, we get 
\begin{equation}
	\cH_\perp = \Omega \tilde{\cH}_\perp,
\hspace{1cm}
	\cH_i = \tilde{\cH}_i.
\end{equation}
On the other hand, since
\begin{eqnarray}
	n_\mu &=& -N (dt)_\mu=-(-g^{00})^{-1/2} (dt)_\mu
	=-(- \Omega^2 \tilg^{00})^{-1/2} (dt)_\mu
\nonumber \\
	&=& \Omega^{-1} \left(- \tilde N (dt)_\mu \right) 
	= \Omega^{-1} \tilde{n}_\mu,
\end{eqnarray}
we have 
\begin{eqnarray}
	\eta^\mu &=& \eta^\perp n^\mu + \eta^i (\partial_i)^\mu
\nonumber \\
	&=& (\eta^\perp \Omega) \tilde{n}^\mu+ \eta^i (\partial_i)^\mu
\\
	&=& \tileta^\perp  \tilde{n}^\mu+ \eta^i (\partial_i)^\mu.
\nonumber
\end{eqnarray}
Hence, we obtain
\begin{equation}
	H[\eta] = \tilde H [\tileta],
\end{equation}
where
\begin{equation}
	\tileta^\perp = \Omega \eta^\perp,
\hspace{2cm}
	\tilde{\eta}{}^i = \eta^i.
\end{equation}

In each conformal frame, the surface term $\cI^i$, which is necessary for well-defined variation principle, has the same {\it form} in terms of quantities in the conformal frame. By a straightforward calculation, we can also show
\begin{equation}
	d\tilde{S}_i~ \tilde{\cI}^i[\tileta]= dS_i~ \cI^i[\eta],
\end{equation}
that is,
\begin{equation}
	\delta \tilde{\cJ}[\tileta] = \delta \cJ[\eta].
\end{equation}
This means that the values of the charges are independent of the choice of conformal frames, up to any terms such as the reference values which vanish for a variation of the canonical variables.


\end{document}